\documentclass[11pt,a4paper,fleqn]{article}
\catcode`\@=11
\@addtoreset{equation}{section}
\def\theequation{\arabic{section}.\arabic{equation}}
\def\appendix{\renewcommand{\thesection}{\Alph{section}}\setcounter{section}{0}
             \renewcommand{\theequation}
           {\mbox{\Alph{section}.\arabic{equation}}}\setcounter{equation}{0}}

\def\maketitle{\thispagestyle{empty}\setcounter{page}0\newpage
               \renewcommand{\thefootnote}{\arabic{footnote}}
                 \setcounter{footnote}0}
\renewcommand{\thanks}[1]{\renewcommand{\thefootnote}{\fnsymbol{footnote}}
              \footnote{#1}\renewcommand{\thefootnote}{\arabic{footnote}}}

\renewcommand{\title}[1]{\begin{center}\Large\bf #1\end{center}\rm\par\bigskip}

\renewcommand{\author}[1]{\begin{center}\Large #1\end{center}}
\newcommand{\address}[1]{\begin{center}\large #1\end{center}}

\def\babs{\hrule\par\begin{description}\item{Abstract: }\it}
\def\eabs{\par\end{description}\hrule\par\medskip\rm}
\renewcommand{\date}[1]{\par\bigskip\par\sl\hfill #1\par\medskip\par\rm}

\newcommand{\dip}[1]{${}^{(#1)}$ Dipartimento di Fisica, Universit\`a di Trento\\
                                     via Sommarive 14, 38123 Trento, Italia\\}
\newcommand{\infn}[1]{${}^{(#1)}\,$TIFPA (INFN), Trento, Italia\\ \medskip}

\newcommand{\sergio}[1]{Sergio Zerbini${}^{#1}$\thanks{e-mail:\sl zerbini@science.unitn.it\rm}}

\def\beq{\begin{eqnarray}}    
\def\be{\begin{eqnarray}}
\def\eeq{\end{eqnarray}}      
\def\ee{\end{eqnarrayn}}

\def\R{{\hbox{{\rm I}\kern-.2em\hbox{\rm R}}}}   
\def\H{{\hbox{{\rm I}\kern-.2em\hbox{\rm H}}}}   
\def\N{{\hbox{{\rm I}\kern-.2em\hbox{\rm N}}}}   
\def\C{{\ \hbox{{\rm I}\kern-.6em\hbox{\bf C}}}} 
\def\Z{{\hbox{{\rm Z}\kern-.4em\hbox{\rm Z}}}}   
\def\dir{/\kern-.7em D\,}                          

\def\be{\begin{equation}}
\def\ee{\end{equation}}
\def\bea{\begin{eqnarray}}
\def\eea{\end{eqnarray}}

\renewcommand{\title}[1]{\begin{center}\Large\bf #1\end{center}\rm\par\bigskip}
\renewcommand{\author}[1]{\begin{center}\Large #1\end{center}}

\begin{document}

\title{ Vacuum Fluctuations as Quantum Probes in FRWL space-times  }  
\author{ \sergio{a,b}}
\address{\dip{a}\infn{b}}

\begin{abstract}

Vacuum fluctuations  related a massless conformally coupled  scalar field in  Friedman-Robertson-Walker-Lemaitre (FRWL) space-times are  investigated. Point-slitting regularization is used and a specific renormalization proposal is discussed. Applications to  generic black holes and FRWL form of the de Sitter space-time are presented.

\end{abstract}


\section{Introduction}

How to probe the properties of the Quantum Vacuum in relativistic quantum field theory (QFT) ? It is well known that this is a difficult task, because one is dealing with relativistic systems having an 
infinite number of degrees of freedom. The possibility we shall discuss here is to  study the vacuum expectation value  $\langle\phi^2(x)\rangle
=<0|\phi^2(x)|0>$ associated with quantum field $\phi(x)$. We will refer to it as  ``vacuum fluctuation''. It contains physical informations on the quantum vacuum, and it is  simpler than the vacuum expectation value of the stress-energy tensor.

In QFT, $\phi(x)$ is an operator valued distribution, thus $\langle\phi^2(x)\rangle$ ill-defined quantity, and  regularization and renormalization procedures are required. 

One may use zeta-function regularization \cite{Eli94,byts}, however, since in cosmology one is dealing with  FRWL space-times and hyperbolic operators,
 point-splitting regularization is  more appropriate.\\ 
Given the  off-diagonal Wightman function
\begin{equation}
 W(x,x') =<0| \phi(x) \phi(x')|0>\,,
\end{equation}
 point-splitting regularization means $x'= x+\varepsilon$, $\varepsilon$ a small cut-off, and the removal of cut-off consists in taking  the  coincidence limit $x'\rightarrow x$.

\section{Fluctuation as Quantum thermometer}

Fluctuations may give informations on thermal nature of the quantum vacuum.
In order to illustrate the issue, let us consider a free massive scalar field defined on the Minkowsky manifold $M^4$,  make a Wick rotation to  imaginary 
compactified time   with period $\beta$. As a result, one is dealing with a massive free quantum  field in thermal equilibrium at temperature  $T=\frac{1}{\beta}$, and  the  relevant operator is defined on $S_1 \times R^3$ and reads
\begin{equation}
 L=-\partial^2_\tau-\nabla^2+M^2\,,
\end{equation} 
$M$ being the mass and $ \tau$  imaginary time with period $\beta$. 
A direct calculation within zeta-function regularization \cite{eli,Cogno} gives for the renormalized fluctuation  associated with this massive free field
\begin{equation}
\langle \phi(x)^2 \rangle_R=\frac{M}{2\pi \beta}\sum_{n=1}^\infty \frac{K_1(n\beta M)}{n}+\frac{M^2}{8\pi^2}\ln\left( \frac{M^2}{\mu^2}\right) \,.
\end{equation} 
If $M \neq 0$,  the thermal properties are not transparent and an arbitrary mass scale $\mu^2$ is present: \\ bad thermometer. 

However,  in the massless case,  there is a drastic simplification and the result is
\begin{equation}
\langle \phi(x)^2 \rangle_R=\frac{1}{12 \beta^2}= \frac{T^2}{12}\,.
\end{equation}
No ambiguity is present, simple reading for the temperature $T$: good thermometer  \cite{Buchholz}.

It is also instructive to present a point-splitting regularization  derivation.  Making use of the images method and KMS condition, it follows that thermal Wigthman 
function  is periodic of period $\beta$ in the imaginary time $\tau$, and reads 
\begin{equation}  
W_\beta(x,x')=\frac{1}{4\pi^2}\sum_n \frac{1}{|\vec{ x}-\vec{x'}|^2+(\tau-\tau'+n\beta)^2}\,.
\end{equation}
The term $n=0$ is the only singular term when $x \rightarrow x'$, and  coincides with the Minkowski contribution.
Renormalization prescription : subtract this term.  Thus,  for  $x \rightarrow x'$, we get again
\begin{equation}  
<\Phi(x)^2>=\frac{1}{2\pi^2\beta^2}\sum_{n=1}^\infty \frac{1}{n^2}=\frac{1}{12 \beta^2}=\frac{T^2}{12}\,.
\end{equation}

\section{Spherically symmetric dynamical black holes}

Fluctuations associated with massless scalar fields  work well as good thermometers in finite temperature QFT in Minkowski space-time. What about the use of fluctuation as thermometers  in presence of gravity? 

Static black holes and  FRWL space-times  in cosmology are  example of  Spherically Symmetric Space-times (SSS), and both  may have an unified descriptions  \cite{noi14}. With regard to this, first let us briefly review the  Kodama-Hayward formalism. A  generic dynamical SSS  may be defined by 
the following  metric  
\begin{equation}
ds^2 =\gamma_{ij}(x^i)dx^idx^j+ R^2(x^i) d\Omega^2\,,\qquad i,j =0,1\;, 
\end{equation}
where the two dimensional {\it normal  metric} is
\begin{equation}
d\gamma^2=\gamma_{ij}(x^i)dx^idx^j\,,
\end{equation}
with  $x^i$  associated coordinates. Furthermore, the quantity $R(x^i)$ is the areal radius, a scalar field in the normal 2-dimensional  metric. 
Dynamical or apparent trapping horizons are   determined by 
\begin{equation}
\gamma^{ij}\partial_iR_H\partial_jR_H=0\,.
\end{equation}
Another important invariant quantity is  the Hayward surface gravity $\kappa_H=\frac{1}{2}(\Delta_\gamma R)_H  $ (see for example \cite{sean,noi09}), which is a generalization of the Killing surface gravity. There is also conserved Kodama vector, $K^i=\frac{\varepsilon^{ij}\partial R_j}{\sqrt{-\gamma}}$, generalization of the Killing vector. Relevant for us are the so called  Kodama observers, defined by condition $R=R_0$.

First Example:  4-dimensional  static  Schwarzschild BH with  $x=(t,r)$ as coordinates in normal space.
\begin{equation}
ds^2=-V(r)dt^2+\frac{dr^2}{V(r)}+r^2 d \Omega^2=d\gamma^2+r^2 d \Omega^2
\end{equation}
with
\begin{equation}
V(r)=1-\frac{2M}{r}\, ,\quad G=1
\end{equation}
Areal radius  $R=r$, the  event horizon $V(r_H)=0$, i.e.  $r_H=2M$,
the surface gravity $\kappa_H=\frac{V'_H}{2}=\frac{1}{4M}$,
Kodama vector is the Killing vector  $K=(1,0,0,0)$. 
Kodama observer: $r=r_0$: constant areal radius.

A  dynamical  SSS example: flat FRWL space-time,  relevant in cosmology
\begin{equation}
ds^2=-dt^2+a^2(t)(dr^2+r^2 d\Omega^2)\,,
\end{equation}
Areal radius is $R=a(t) r$, physical radial distance, $H(t)=\frac{\dot{a}}{a}$ is the Hubble parameter.
There is a  dynamical trapping horizon (Hubble radius)
\begin{equation}
R_H=\frac{1}{H(t)}\,.
\end{equation}
The Kodama vector is 
\begin{equation}
K=(1, -H(t)R,0,0)
\end{equation}
Kodama observer: $R=R_0$ constant areal radius, namely $r=\frac{R_0}{a(t)}$.

\section{Fluctuations in FRWL space-times  }
The main idea: use  the fluctuation related to a  suitable quantum probe, which has to be a good thermometer. 
The probe: a conformally coupled massless scalar  field.  Its Lagrangian reads
\begin{equation}
L=\sqrt{-g}\left(-\frac{1}{2}\partial^i \phi \partial_i \phi-\frac{1}{6}\mathcal R \phi^2 \right)\,,
\end{equation} 
$\mathcal R$ Ricci scalar curvature. 

It is convenient to re-write the flat FRWL  space-time making us of  the conformal time $\eta$,  $d\eta=\frac{d t}{a}$,  
\begin{equation}
ds^2=-dt^2+a^2(t)d \vec x^2=a^2(\eta)(-d\eta^2+d \vec x^2)\,.
\end{equation}

Recall that the  Wightmann function is $W(x,x')=<0|\phi(x)\phi(x')|0>$. Now   the  field $\phi(x)$ is ``free'' and admits the expansion  
\begin{equation}
\phi(x)=\sum_{\vec k}f_{\vec k}(x)a_{\vec k}+h.c.\,.  
\end{equation}
where $f_{\vec k}(x)$ are the associated modes. The related Conformal Vacuum  is defined by 
\begin{equation}
a_{\vec k }|0>=0  \,.
\end{equation}
As a result
\begin{equation}
W(x,x')=\sum_{\vec k} f_{\vec k}(x)f^*_{\vec k}(x')\,,
\end{equation}
where the modes functions $f_{\vec k}(x)$ satisfy  conformally invariant equations
\begin{equation}
\left(\Delta -\frac{\mathcal R}{6}\right) f_{\vec k}(x)=0\;.
\end{equation}
In this case, solutions of this equation are simple and  characterize  the Conformal Vacuum:
\begin{equation}
f_{\vec k}(x)= \frac{e^{-i\eta k }}{2\sqrt{ k}a(\eta)}  e^{-i\vec k \cdot \vec x}\,\quad k= |\vec k |
\end{equation}

The related Wightmann function  $W(x,x')$ can be evaluated, and this covariant bi-scalar distribution  reads
\begin{equation}
W(x,x')= \frac{1}{4\pi^2 a(\eta)a(\eta')}\,\frac{1}{|\vec x- \vec x'|^2-|\eta-\eta'-i\epsilon|^2 }\,.
\end{equation}  
 It is convenient to make use of the proper time $\tau$ as evolution parameter. As a result,  
the proper-time-parametrized Wightman function is
\begin{equation}
 W(x(\tau),x'(\tau')) = \frac{1}{4 \pi^2} \frac{1}{\sigma^2(\tau, \tau')}\, ,
\end{equation}
where the invariant distance is defined by
\begin{equation}
 \sigma^2(\tau,\tau') = a(\tau) a(\tau')\, \Big( x(\tau) - x(\tau') \Big)^2\,.
\end{equation}
 
\section{Point-splitting regularization}
Due to the isotropy of flat  FRWL space-time, one may restrict only to radial time-like  trajectories, namely $x(\tau)= \big(\eta(\tau),r(\tau)\big)$.  

Putting $\tau'=\tau +\varepsilon $, with  $\varepsilon $ small and $\dot{ t}=\frac{d t}{d \tau}$, and introducing  the four-acceleration along the trajectory from $x$ to $x' $,  
\begin{equation}
 A^2 = \left[ \frac{\ddot{t}}{\sqrt{\dot{t}^2-1}} +  H \sqrt{ \dot{t}^2-1}  \right]^2\,,
\end{equation}
one obtains
\begin{equation}
 \sigma^2(\tau, \varepsilon ) = - \varepsilon^2 - \frac{1}{12} \Big[ A^2 + H^2 + 2 \dot{t}^2 \partial_t H \Big] \varepsilon^4 + O(\varepsilon^6)\,.
\end{equation}
As a result, for small   $\varepsilon$  the Wightman function is
\begin{equation}
W(\tau, \varepsilon)=-\frac{1}{4\pi^2}\frac{1}{\varepsilon^2}+ \frac{1}{48 \pi^2} \Big[ A^2 + H^2 + 2 \dot{t}^2 \partial_t H \Big]+O(\varepsilon^2)\,.
\end{equation}
\section{Renormalization}

In Minkowski space-time, one has $H(t)=0$, and for inertial trajectories, one has  vanishing acceleration, namely
\begin{equation}
 W_{M}(\tau, \varepsilon ) = -\frac{1}{4\pi^2}\frac{1}{\varepsilon^2}  \,. 
\end{equation}
Choice of the renormalization prescription:  subtract  this contribution. As a consequence,  the renormalized vacuum fluctuation  reads \cite{noi14}  
\begin{equation}
\langle \phi^2(x) \rangle_R  = \frac{1}{48 \pi^2} \left[   A^2 + H^2 + 2 \dot{t}^2 \partial_t H   \right]\,,  
\label{r}
\end{equation}
Some remarks are in order.   
The first term is  depending on the  trajectory through $A$ and hence $\dot{t}$. The second  one is depending on the dynamical space-time through $H$, 
 and the third term is an ``entangled'' contribution, vanishing for stationary space-times.

Important byproduct. In Minkowski case, $H=0$. Thus one has only two cases:

If  $A$ not vanishing: one has  the Unruh effect, and the related  Unruh temperature is 
\begin{equation}
T_U=\frac{A}{2\pi}\,. 
\end{equation}
On the other hand, if $A=0$, namely inertial observer, one has   
 $\langle \phi^2(x) \rangle_R=0$, and  the  Minkowski renormalized result is obtained. 

\section{de Sitter FRWL space-time}

A very important example of FRWL space-time is  de Sitter one, its flat FRWL metric being
\begin{equation} 
ds^2=-dt^2+e^{2H_0t} d \vec x^2\,,
\end{equation}
 $a(t)=e^{H_0t}$, $H(t)=H_0$  constant.

For Kodama observers with $R=R_0$, the conformal time can be evaluated in terms of proper time $\tau$
\begin{equation}
\eta(\tau)=\frac{1}{H_0}e^{-\frac{H_0}{\sqrt{1-R_0^2 H_0^2}}\tau}
\end{equation}
and the dS invariant  distance is 
\begin{equation}
\sigma^2(s)=-\frac{(1-R_0^2H_0^2)}{4 H_0^2} \sinh^2\left( \frac{H_0}{2\sqrt{1-R_0^2 H_0^2}}s\right)
\end{equation}
This is an example of  stationary space-time: it depends only on the difference  $s=\tau-\tau' $.

General formula (\ref{r}) leads to  the fluctuation for dS 

\begin{equation}
\langle \phi^2(x) \rangle_R =\frac{1}{48\pi^2}\left(\frac{H_0^4R_0^2}{(1-R_0^2 H_0^2)}+H_0^2 \right)\,,
\end{equation}
the first term in the bracket being the acceleration of the Kodama observer at fixed $R_0$. One also has
\begin{equation}
\langle \phi^2(x) \rangle_R =\frac{1}{12}\frac{T_{GH}^2}{(1-R_0^2 H_0^2)}\,.
\end{equation}
  $T_{GH}=\frac{H_0}{2\pi}$ is the well known Gibbons-Hawking temperature,
 $\frac{1}{(1-R_0^2 H_0^2)}$  being a kinematical red-shift factor. One is dealing with a   good thermometer.

\section{de Sitter space as  black holes}
 The de Sitter space-time admits also  static patch  
\begin{equation}
ds^2=- (1-H^2_0 r^2)  dt_s^2+\frac{dr^2}{1-H^2_0 r^2 }+r^2 d \Omega^2 \,,
\end{equation}
 $t_s$ is the time coordinate, and the areal radius is $R=r$.
The related  horizon: $r_H=\frac{1}{H_0}$, and  surface gravity $\kappa_H=H_0$. \\
In the cosmological dS patch, the fluctuation leads, modolo a red-shift factor,  to the Gibbons-Hawking temperature $T_H=\frac{H_0}{2\pi}$. 
What about the temperature issue in this static patch? 
One can answer to this question in general.

\section{Static black hole as effective FRWL space-time}
Black holes are not black, and it is well known that they emit a quantum Hawking radiation in thermodynamical equilibrium at temperature $T_H$. 
In order to (partially) investigate this fundamental issue, first let us show that for Kodama observers, one is dealing with a special FRWL space-time.
Start with a generic static BH solution 
 \begin{equation}
ds^2=-V(r)dt_s^2+\frac{dr^2}{V(r)}+r^2 d \Omega^2 \nonumber 
\end{equation}
with event horizon at $V(r_H)=0$, $V'_H \neq 0$. 
In order to avoid the metric singularity at $r=r_H$, one can make use of Kruskal coordinates. 
With tortoise radial coordinate $dr^*=\frac{dr}{V(r)}$, the BH Kruskal metric is
\begin{equation}
ds^2 = e^{-2\kappa_H r^*}\,V(r^*)[-dT^2+dX^2]+r^2(T,R) d \Omega^2\,, 
\end{equation}
where now $r^*=r^*(T,X)$, $\kappa_H=\frac{V'_H}{2}$  being the Killing surface gravity.
The relevant part of BH Kruskal metric is its normal part, namely
\begin{equation}
d\gamma^2 = e^{-2\kappa_H r^*}\,V(r^*)[-dT^2+dX^2]\,. 
\end{equation}
First key point : Kruskal normal space-time is conformally related  to two-dimensional Minkowski space-time.
 
Second point:  Kruskal  space-time is  effective flat  FRWL space-time for  Kodama observers  $r=r_0$
\begin{equation}
d\gamma^2=V_0e^{-2\kappa_H r_0^*}(-dT^2+dX^2)\,,
\end{equation}
and  we also have 
\begin{equation}
d\gamma^2=-dt^2+a_*^2 dX^2=-dt^2_sV_0\,.
\end{equation}
Thus  $t=\sqrt{V_0}\,e^{-\kappa_H r_0^*}\,T$ is a  ``cosmological''  time, 
and $a(r_0^*)=\sqrt{V_0}e^{-\kappa_H r_0^*}$ is the related constant ``expansion" factor, and $d\tau^2=dt^2_s\,V_0$.

Since  $a^2(r_0)$ constant, i.e. $  H=0$, only acceleration term is present, and the general formula (\ref{r}) gives
\begin{equation}
\langle \phi^2(x) \rangle_R  = \frac{1}{48 \pi^2} \left[ \frac{(\ddot{t})^2}{\dot{t}^2-1} \right]\,. 
\end{equation}
Using Kruskal coordinates definition and $t_s=\frac{\tau}{\sqrt{V_0}}$, one gets
\begin{equation}
\dot{t}  =  \cosh\left(\kappa_H\frac{\tau}{\sqrt{V_0}}\right)\,, \quad 
\ddot{t}= \frac{\kappa_H}{\sqrt{V_0}}\sinh\left(\kappa_H\frac{\tau}{\sqrt{V_0}}\right)\,.
\end{equation}
As a result 
\begin{equation}
\langle \phi^2(x) \rangle_R  = \frac{1}{48 \pi^2}  \frac{\kappa_H^2}{V_0}=\frac{1}{12}\,\frac{T_H^2}{ V_0}\,. 
\end{equation}
Again a good  thermometer: local temperature is
\begin{equation}
T_0=\frac{T_H}{\sqrt{V_0}}\,,
\end{equation}
with the Hawking temperature  $T_H=\frac{\kappa_H}{2\pi}$, while
 $V_0=-g_{00}$ being the Tolman red-shift factor.\\
In the  de Sitter space-time,  $V(r)=1-H_0^2r^2$ and  $T_H=\frac{|\kappa_H|}{2\pi}=\frac{H_0}{2\pi }$, 
in full agreement with previous result.

\section{Conclusion}
Fluctuation related to  massless conformally coupled scalar field has been used to probe the Quantum Vacuum in presence of gravity.
Making use of this  quantum probe, after a point-splitting regularization in the proper-time, the following  renormalization prescription has been used: remove the Minkowski contribution related to  inertial observers. In principle, other renormalization prescriptions are possible. However, one can show that our renormalization prescription is full agreement with the results obtained making use of Unruh- de Witt detector approach \cite{noi11}.

\noindent {\bf Acknowledgements.} 

I thanks L. Vanzo, L. Bonetti and G. Acquaviva for useful discussions.
This research has been supported by the  INFN grant, project FLAG 2014.



\begin{thebibliography}{99}




\bibitem{Eli94} 
  E.~Elizalde, S.~D.~Odintsov, A.~Romeo, A.~A.~Bytsenko and S.~Zerbini,
  ``Zeta regularization techniques with applications,''
  Singapore, Singapore: World Scientific (1994) 319 p.

\bibitem{byts} 
  A.~A.~Bytsenko, G.~Cognola, E.~Elizalde, V.~Moretti and S.~Zerbini,
  ``Analytic aspects of quantum fields,''
  River Edge, USA: World Scientific (2003) 350 p.

\bibitem{eli} 
  G.~Cognola, E.~Elizalde and S.~Zerbini,
  Phys.\ Rev.\ D {\bf 65}, 085031 (2002)
  [hep-th/0201152].


\bibitem{Cogno} 
  G.~Cognola and S.~Zerbini,
  Int.\ J.\ Mod.\ Phys.\ A {\bf 18}, 2067 (2003).





\bibitem{Buchholz} 
  D.~Buchholz and J.~Schlemmer,
  Class.\ Quant.\ Grav.\  {\bf 24}, F25 (2007)
  [gr-qc/0608133].


\bibitem{noi14} 
  Acquaviva G., Bonetti L., Vanzo L.  and Zerbini S. (2014)
  {\it Phys.\ Rev.\ D} {\bf 89}, 084031.
 



\bibitem{sean} 
  R.~Di Criscienzo, S.~A.~Hayward, M.~Nadalini, L.~Vanzo and S.~Zerbini,
  Class.\ Quant.\ Grav.\  {\bf 27}, 015006 (2010).





\bibitem{noi09}

Hayward S. A., Di Criscienzo R., Vanzo L., Nadalini M. and S.~Zerbini S. (2009)
  {\it Class.\ Quant.\ Grav.}  {\bf 26}, 062001. 
  [arXiv:0806.0014 [gr-qc]].

\bibitem{noi11} 
  Acquaviva G., Di Criscienzo R., Tolotti M., Vanzo L. and Zerbini S. (2012)
 {\it Int.\ J.\ Theor.\ Phys.}  {\bf 51}, 1555 
  [arXiv:1111.6389 [gr-qc]].





\end{thebibliography}
\end{document}